# Charge-Density-Wave like Behavior in the One-Dimensional Charge-Ordered Semiconductor $(NbSe_4)_3I$

D. Starešinic[1,2], P. Lunkenheimer[1], J. Hemberger[1], K. Biljakovic[2], and A. Loidl[1]

[1] Experimental Physics V, Center for Electronic Correlations and Magnetism, University of Augsburg, D-86135 Augsburg, Germany
[2] Institute of Physics, University of Zagreb, HR-10001 Zagreb, Croatia

We report on broadband dielectric spectroscopy on the one-dimensional semiconductor $(NbSe_4)_3I$. Below the structural phase transition close to 270 K we observe colossal dielectric constants with a frequency and temperature dependence very similar to what is observed in canonical charge-density wave systems. Guided by structural details we interpret this structural phase transition as driven by complex charge-order processes.

One-dimensional (1D) systems are fascinating topics of recent modern solid state physics, both theoretically and experimentally. 1D magnets and 1D metals are well studied species and a number of experimental realizations are known. More than 50 years ago it was demonstrated by Peierls and Fröhlich that a one-dimensional electron gas coupled to phonons is unstable and undergoes a metal-to insulator transition. This new type of ground-state condensate, a charge-density wave (CDW) has first been observed in blue bronze. Iodine doped transition-metal tetraselenides $(MSe_4)_nI$ with M = Nb, Ta and n = 2, 3, and 10/3, turned out to be paramount examples of materials with one-dimensional electronic structure exhibiting different band fillings [1,2,3,4,5]. For example, $(TaSe_4)_2I$, $(NbSe_4)_2I$, and $(NbSe_4)_{10/3}I$ are metallic at room temperature and undergo classical Peierls transitions with CDW formation at 263 K [6, 7], 220 K [7] and 285 K [8] respectively. However, $(NbSe_4)_3I$ was found to be a charge-ordered semiconductor at room temperature, exhibiting a displacive structural phase transition at 274 K [9] to another insulating phase not associated with CDW formation. Experimentally, 1D insulators can only be identified via hopping processes of charge carriers and certainly also via the band structure. Indeed, from optical spectroscopy and photoemission, $(NbSe_4)_3I$ has been characterized as 1D insulator with characteristic dynamic properties [10].

It is well known that at audio frequencies CDW systems reveal unusually large dielectric constants and strong dispersion effects as discussed in detail by Cava et al. [11,12] and as observed also for the systems $(TaSe_4)_2I$ and $(NbSe_4)_{10/3}I$ [13]. This low-frequency dielectric behavior in CDW systems was explained by Littlewood [14] in terms of screening effects of the pinned CDW. Hence, in CDW systems dielectric spectroscopy is a supreme technique yielding important and characteristic results for the pinned charge density wave. Naively, the formation of CDWs can be viewed as a specific type of charge order (CO). In transition-metal oxides, CO is a common phenomenon and an increasing number of dielectric measurements close to charge-ordering transitions are reported, specifically in low-dimensional systems having in mind the close analogy to the Peierls transition in metallic systems. Indeed, as mentioned above, a giant dielectric response has been detected in the insulating two-leg ladder $Sr_{14}Cu_{24}O_{41}$ [15,16] and in one-dimensional $(DI-DCNQ)_2Ag$ [17]. Thus it seems promising to investigate also the insulating charge-ordered tetraselenide $(NbSe_4)_3I$ by broadband dielectric spectroscopy. Here we report the dielectric permittivity from sub-audio to radiowave frequencies. Below the structural phase transition at 270 K, we find frequency and temperature dependent dielectric constants, very similar to what is observed in canonical charge-density wave systems [11,12], despite the fact that the phase transition in $(NbSe_4)_3I$ can not be characterized as a Peierls instability.

Single crystalline samples of $(NbSe_4)_3I$ were prepared by vapor transport. The samples have been characterized by dc resistivity, heat capacity and magnetization measurements, which will be reported separately [18]. The dc resistivity at room temperature is of the order 1.6 Ωcm, close to the values reported in literature [9]. From the temperature dependence of the heat capacity we deduce a structural phase transition temperature of $T_s = 268 \pm 1$ K. The broadband dielectric response from 10 mHz to 100 MHz has been determined utilizing frequency-response analysis and a reflectometric technique [19] employing the Novocontrol α-analyzer and the impedance analyzer Agilent 4294A. Measurements on needle-shaped samples were performed along the crystallographic c-direction for temperatures from 1.5 K < T < 300 K. Special care has been taken to determine the true intrinsic dielectric response and to rule out contact or surface-layer contributions, which may play a role in dielectric measurements usually performed in two-point contact configuration [20]. This has been done by performing measurements on samples with different contacts and with very different (up to a factor of ten) lengths.

Fig. 1(a) shows the temperature dependence of the real part of the dielectric constant ε'. Starting at room temperature, the dielectric constant increases, reaches colossal values of the order of $10^5$ between 100 and 200 K, and reveals a strong drop to values below 100 towards low temperatures. From our detailed measurements with different contacts and different samples geometries we conclude that the dispersion effects for T > 200 K which appear at high frequencies are driven by contact contribution. However, below 200 K all results represent the purely intrinsic dielectric response of $(NbSe_4)_3I$. The millimeter-wave results of ref. [21], measured contact-free utilizing quasi-optic spectroscopy, match well into the observed frequency dependence of Fig. 1(a). Astonishingly, the temperature dependence of the dielectric constant closely resembles the observations in the CDW systems $(NbSe_4)_{10/3}I$ and $(TaSe_4)_2I$. In these compounds the dielectric constant strongly increases in the Peierls state reaching values of $10^6$ and smoothly decreases towards low temperatures, again revealing strong dispersion effects [13].

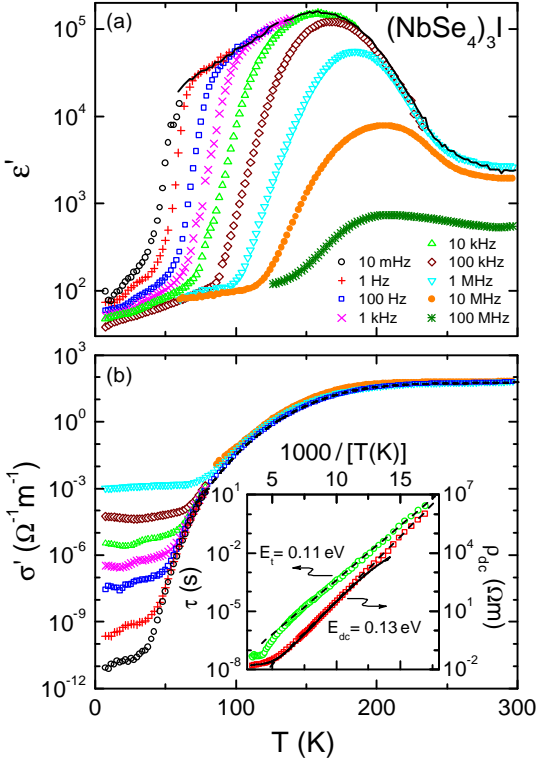

FIG. 1 (color online). Temperature dependence of the dielectric constant (a) and conductivity (b) for various frequencies. The line in (a) represents the static susceptibility as determined from an analysis of the frequency-dependent data (Fig. 2). The line in (b) shows the dc conductivity measured by a standard 4-point technique. The inset shows the relaxation time (circles; from an analysis of the frequency-dependent spectra, Fig. 2) and the dc conductivity [solid line: 4-point measurement, squares: from $\sigma'(\nu)$] in an Arrhenius representation. The dashed lines demonstrate thermally activated behavior.

Fig. 1(b) shows the conductivity, $\sigma' = \omega\varepsilon_0\varepsilon''$, where the angular frequency $\omega = 2\pi\nu$, $\varepsilon_0$ is the permittivity of free space and $\varepsilon''$ is the dielectric loss. Frequency independent dc-conductivity dominates for $T > 100$ K, while ac hopping conductivity comes into play at lower temperatures. The dashed lines indicate the result of a 4-point dc-conductivity measurement, performed down to 70 K. It precisely agrees with $\sigma'$ measured at 10 mHz, giving further evidence that contact effects play no role. In the inset, the dc-conductivity determined from the dielectric and the 4-point measurements is plotted in an Arrhenius representation (squares). The dashed line indicates that the dc conductivity follows thermally activated behavior with a hindering barrier of $0.13$ eV $\approx 1500$ K. Comparing the conductivity of the CO "insulating" tetraselenide ($n = 3$) with the "metallic" high temperature phases of the CDW systems with $n = 2$ and $n = 10/3$, we find that the difference amounts two orders in magnitude at most. In $(NbSe_4)_3I$ the room-temperature resistivity is about $1.6$ $\Omega$cm in good agreement with published results [9], while in $(TaSe_4)_2I$ the resistivity is of the order $1.5 - 3\times10^{-3}$ $\Omega$cm [6] and in $(NbSe_4)_{10/3}I$ $1.5\times10^{-2}$ $\Omega$cm [8]. It is interesting that at room temperature all tetraselenides exhibit a temperature dependence of the resistivity which rather signals semiconducting behavior. Even in the insulating, charge-ordered title compound there obviously is considerable electron transfer along the chains.

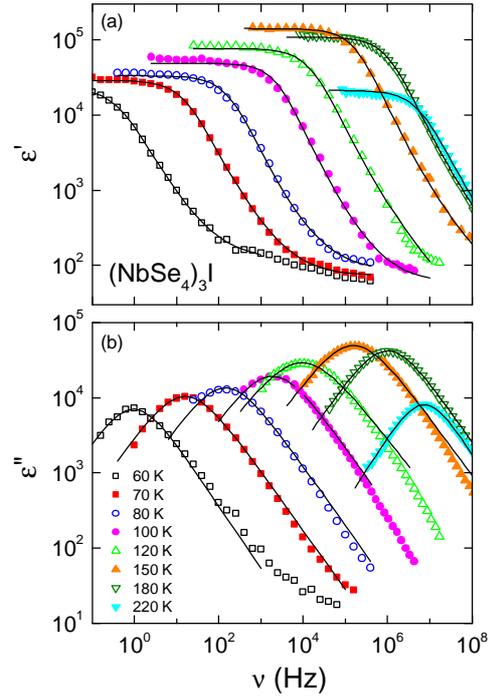

FIG. 2 (color online). Frequency-dependence of the dielectric constant (a) and loss (b) for various temperatures. The lines are fits with the empirical Cole-Cole function.

Fig. 2 documents the frequency dependence of the dielectric constant (a) and of the dielectric loss (b) for various temperatures between 60 K and 220 K. The dielectric constant shows a step-like decrease on increasing frequency, which is typical for relaxational behavior. The point of inflection strongly shifts towards lower frequencies with decreasing temperature, indicating a slowing-down of the mean relaxation rate. To get an estimate of the relaxational part of the dielectric loss, all ac and dc conductivities have been subtracted from the raw data, yielding well defined loss peaks, which also shift through the frequency window on decreasing temperature. Real and imaginary part of the dielectric constant can be consistently described with the phenomenological Cole-Cole equation, which describes a symmetrically broadened Debye-relaxation utilizing an extra width parameter $\alpha$ between 1 (Debye case) and 0 (infinitely broad). The broadening commonly is ascribed to a distribution of relaxation times. The results of the simultaneous fits to real and imaginary part of the dielectric permittivity are indicated as solid lines in Fig. 2. The width parameter for all temperatures was close to 0.8, indicating a moderate distribution of relaxation times. The temperature dependence of the static dipolar susceptibility, determined by the height of the step of ε' and by the area under the loss peaks ε", is indicated as solid line in Fig. 1(a). It closely follows the temperature dependence of the dielectric constant measured at 10 mHz down to 50 K. Finally the temperature dependence of the mean relaxation time, $\tau = 1/(2\pi\nu_p)$, with $\nu_p$ the maximum of the loss peaks, is shown in the inset of Fig. 1(b) (circles). Here we used an Arrhenius type of presentation, which leads to a linear dependence of the mean relaxation time between 60 K and 150 K (dashed line), resulting in an energy barrier against thermal dipolar relaxations of 0.11 eV (≈ 1300 K). This value is only slightly lower than the hindering barrier of 0.13 eV determined from the dc conductivity, which is plotted in the same frame.

The occurrence of a very strong relaxational mode in $(NbSe_4)_3I$, as revealed in Figs. 1 and 2, is highly unexpected and an interpretation is not straight forward. To get an insight into the possible reason for this unusual behavior, one has to consider the crystalline structure of the tetraselenides: They crystallize in tetragonal structures with the transition metal arranged along the c-axis separated by rectangular $Se_4$ units [schematically shown in Fig. 3(a)]. Two adjacent $Se_4$ rectangles are twisted by a well defined angle depending on the specific compound under consideration and in the tetragonal structure these $MSe_4$ chains are well separated by iodine ions. In all tetraselenides the Nb $d_{z^2}$ orbitals, which show overlap along c, are partly filled, e. g. with a filling of 1/4 for n =2 and 1/3 for n =3 and hence are expected to be metallic at high temperatures. This metallic phase is unstable against a Peierls distortion yielding a CDW state for n = 2 and n = 10/3. However, $(NbSe_4)_3I$ reveals a different ground state with a specific type of CO. This compound can be written as $Nb_2^{4+}Nb^{5+}(Se_2^{2-})_6I^-$ [5], and hence along the c axis a CO with a characteristic sequence of filled and empty d-orbitals, namely $Nb^{4+}$ ($d^1$), $Nb^{4+}(d^1)$, and $Nb^{5+}(d^0)$ with concomitant orbital order is realized [Fig. 3(b)]. This prominent CO and orbital arrangement, with strong covalent bonds between $d^1$ ions and weak bonds between $d^1$ and $d^0$ niobium ions comes along with very different bond-lengths along c. The crystallographic structure with six Nb ions per chain and per lattice spacing exhibits a sequence of long (L: 0.325 nm) and short (S: 0.306 nm) bond lengths, namely LLSLLS, with the short bonds between the $Nb^{4+}$ ions [Figs. 3(a) and (b)]. Obviously the $d^1$ orbitals exhibit an orbital-derived dimerisation with $d_{z^2}$ - $d_{z^2}$ dimers characterized by a short bond length and by a spin singlet. These $Nb^{4+}$ dimers are separated by single $Nb^{5+}$ ions without any orbital and spin degrees of freedom. At the structural phase transition close to 270 K the $Nb^{5+}$ becomes attached closer to one dimer and the sequence of bond lengths now becomes LISLIS, with an additional intermediate (I) bond length (I: 0.317nm), a slightly increased bond L (0.331), and S (0.306nm) remaining almost unchanged [Fig. 3(c)] [5]. When compared to the L bonds in the high-temperature phase the new bond L is increased by 0.006 nm and the bond I is decreased by 0.008 nm. From this observation it is clear, that in the high-temperature phase the $Nb^{5+}$ ion is located in a double-well potential, on cooling selecting one side of this double-well potential as a stable equilibrium position, which explains the occurrence of a displacive second order phase transition. At the same time two neighboring $NbSe_4$ chains are shifted against each other [5]. The overall situation is sketched in Fig. 3 (based on the assumption of truly localized electrons which certainly is oversimplified).

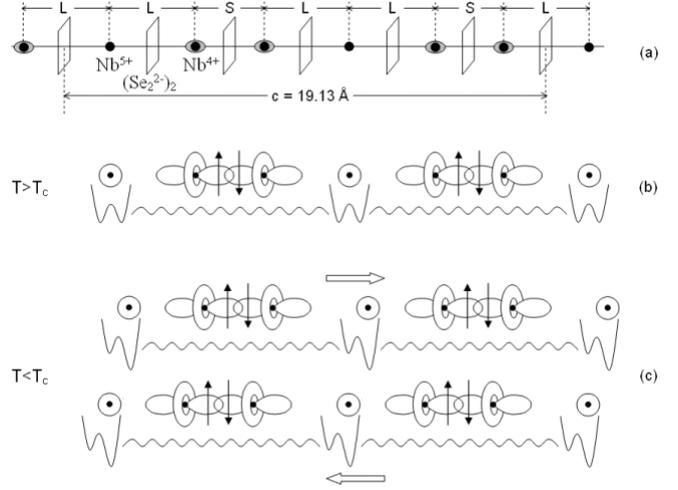

FIG. 3. Schematic drawing of the chain structure of $(NbSe_4)_3I$ above [(a) and (b)] and below (c) the structural phase transition temperature $T_c$ = 170 K.

One now may speculate that the observed giant dielectric response in $(NbSe_4)_3I$ is related to the displacive motion of the $Nb^{5+}$ ion within its double-well potential. This corresponds to the reorientation of a relatively strong dipolar moment and thus may lead to a strong relaxation feature. This motion of the $Nb^{5+}$ ions must be highly collective as it leads to the phase transition at 270 K. Fig. 3(b) also implies that $(NbSe_4)_3I$ may be a good realization of a one-dimensional ferroelectric, when neglecting the shifts of neighboring chains with respect to each other. In this context, it is interesting that in many respects, the dielectric response of $(NbSe_4)_3I$ resembles the observations in relaxor ferroelectrics, namely on decreasing temperatures a Curie-Weiss like increase, a broad smeared out maximum and strong dispersion effects on the low-temperature side of the cusp-shaped maximum [22] [Fig. 1(a)]. Relaxor behavior is observed in diluted ferroelectrics where long-range ferroelectric order is suppressed by disorder and frustration. The tetraselenide under investigation is a purely stoichiometric compound and long-range polar order may be suppressed via dimensionality, e. g. like in one-dimensional magnets, which are characterized by a broad smeared out maximum in the magnetic susceptibility at a temperature corresponding to the average magnetic exchange interaction. However, the low-temperature structure of $(NbSe_4)_3I$ has been characterized as non-centrosymmetric but with no unique polar axis, so that from a 3D point of view ferroelectricity can not be expected [5]. In addition, we have to state that we were not able to detect any pyro-currents or ferroelectric hysteresis effects, even not at the lowest temperatures, which, however, may be due to a suppression of these features by conductivity contributions. As the dielectric response of (NbSe$_4$)$_3$I closely resembles that of CDW systems, and especially that of the closely related tetraselenides (TaSe$_4$)$_2$I and (NbSe$_4$)$_{10/3}$I [13], it is straightforward to alternatively consider an explanation similar to that in CDW systems. The CO in (NbSe$_4$)$_3$I could be regarded as a special kind of (commensurate) CDW and it may be possible that screening effects by free charge carriers play a role.

Finally, irrespective of the microscopic origin of the detected dielectric behavior, it is notable that (NbSe$_4$)$_3$I represents a semiconducting material with extremely large ("colossal") dielectric constants of truly intrinsic nature. The preparation of a usable material with large dielectric constants (> 1000, preferably in a large frequency window and in a broad temperature range) would be a milestone for modern electronics and thus recent reports on the appearance of colossal dielectric constants in various types of transition-metal oxides [23,24,25] gained considerable interest. However, meanwhile it became clear that some of these CDCs are only apparent and can be explained assuming Schottky-type depletion layers or thin insulating layers at the boundaries between bulk sample and contacts [20,26]. Nevertheless the race for materials with CDCs will continue and the present results corroborate the experimental evidence that charge-ordered materials are good candidates.

In conclusion, dielectric spectroscopy on the charge-ordered 1D tetraselenide (NbSe$_4$)$_3$I revealed a strong relaxation mode and the occurrence of CDCs, which closely resembles the typical findings in CDW systems, and especially in the structurally related tetraselenides (TaSe$_4$)$_2$I and (NbSe$_4$)$_{10/3}$I [13]. Considering the crystalline structure of this compound, an explanation in terms of ferroelectrically-correlated displacive ionic motions seems possible. However, having in mind the mentioned close resemblance to CDW systems, it seems reasonable to alternatively consider an explanation similar to that in CDW systems. The CO in (NbSe$_4$)$_3$I can be regarded as a special kind of (commensurate) CDW and it may be possible that screening effects by free charge carriers play a role in the generation of the observed giant relaxation mode. While our results reveal a most spectacular and unexpected dielectric behavior of this compound, dielectric spectroscopy alone certainly cannot resolve the question of the true microscopic origin of this behavior and thus further investigations of (NbSe$_4$)$_3$I are necessary.

We thank F. Levy for providing high quality single crystals. This work was partly supported by the Deutsche Forschungsgemeinschaft via the Sonderforschungsbereich 484 and partly by the BMBF via VDI/EKM.